\documentclass{aa}
\usepackage{epsf}

\begin{document}

\title{The Cooling Neutron Star in 3C 58}
\author{
D.~G. Yakovlev \inst{1}
\and
A.~D. Kaminker \inst{1}
\and
P. Haensel \inst{2}
\and
O.~Y. Gnedin \inst{3}
}
\institute{
Ioffe Physical Technical Institute,
         Politekhnicheskaya 26, 194021 St.~Petersburg, Russia
\and
N.\ Copernicus Astronomical Center,
         Bartycka 18, 00-716 Warsaw, Poland
\and
Space Telescope Science Institute,
         3700 San Martin Drive, Baltimore, MD 21218, USA
\\
{\em  yak@astro.ioffe.rssi.ru,
kam@astro.ioffe.rssi.ru,
haensel@camk.edu.pl,
ognedin@stsci.edu
}}
\offprints{D.G.\ Yakovlev}

\date{Received x xxx 2002 / Accepted x xxx 2002}
\abstract{
The upper limit on the effective surface
temperature of the neutron star (NS) PSR J0205+6449
in the supernova remnant 3C 58 obtained recently by Slane
et al.\ (\cite{shm02}) is analyzed using
a modern theory of NS cooling (Kaminker et al.\ \cite{kyg02}).
The observations can be explained by
cooling of a superfluid NS with the core composed of
neutrons, protons, and electrons, where direct Urca process
is forbidden. 
However, combined with the data on the surface temperatures
of other isolated NSs, it gives evidence 
(emphasized by Slane et al.) that direct
Urca process is open in the inner cores of massive NSs.
This evidence turns out to be 
less stringent than that provided by
the well known observations of Vela and Geminga.
}
\titlerunning{The cooling neutron star in 3C 58}
\authorrunning{D.~G.\ Yakovlev, A.~D.\ Kaminker, P.\ Haensel, O.~Y.\ Gnedin}
\maketitle

\section{Introduction}
\label{sect-intro}

PSR J0205+6449, a pulsating X-ray source, was
discovered by Murray et al.\ (\cite{murrayetal02})
in the supernova remnant 3C 58, 
which is most likely associated
with the historical supernova SN 1181. It is thus one of the youngest
neutron stars (NSs) observed. 
Recently, using {\it Chandra} observations Slane et al.\ (\cite{shm02})
inferred an upper limit on the
effective surface temperature (redshifted
for a distant observer):
$T_{\rm s}^\infty < 1.08$ MK. 
They emphasized that
this upper limit for a young NS
is low,
``suggesting the presence of some exotic cooling contribution
in the interior''.
In other words,
it provides evidence for the powerful direct Urca
process (Lattimer et al.\ \cite{lpph91}) 
in the NS inner core, 
or for similar processes of enhanced neutrino emission
in pion-condensed, kaon-condensed, 
or quark core, as reviewed, e.g., by Pethick
(\cite{pethick92}).

Here we analyze this intriguing possibility
in more detail using recent results of the NS cooling
theory (Kaminker et al.\ \cite{khy01};  
Yakovlev et al.\ \cite{ykg01};
Kaminker et al.\ \cite{kyg02}, hereafter KYG;
Yakovlev et al.\ \cite{ygkp02}, hereafter YGKP)
and taking into account the observational data on thermal
emission of other isolated middle-aged NSs.  

The observational basis is shown in Figs.\ 1 and 2.
They display the upper limit of $T_{\rm s}^\infty$
for PSR J0205+6449 with the age of SN 1181,
and the observational values of $T_{\rm s}^\infty$
for eight middle-aged isolated NSs, the same as in KYG and YGKP.
They are:
RX J0822--43, 
1E 1207--52, 
and RX J0002+62 
(radio-quiet NSs in supernova remnants);
Vela, 
PSR 0656+14, 
Geminga, 
and PSR 1055--52 
(observed as radiopulsars);
and RX J1856--3754 
(also a radio-quiet NS).
We do not analyze the less likely possibility that
the age of J0205+6449 is given by the pulsar
dynamical age $\approx 5400$ yr,
measured by Murray et al.\ (\cite{murrayetal02});
that case could be easier explained by the 
cooling theory.
The values of
$T_{\rm s}^\infty$
and $t$ for other sources are the same as in KYG and YGKP,
with the only exception of the age
of RX J1856--3754, $t=5 \times 10^5$ yr, as revised
recently by Walter \& Lattimer (\cite{wl02}). 
Note also a too slow spindown rate
of 1E 1207--52 measured by Pavlov et al.\ (\cite{pzst02})
which may cast doubts on the correct determination of the age of this NS.

\begin{figure}
\centering
\epsfxsize=86mm
\epsffile[20 143 565 690]{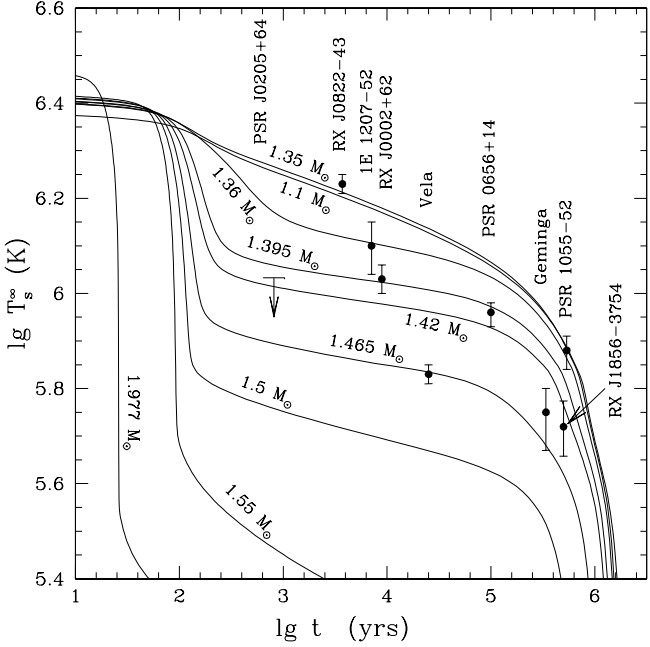}
\caption{
Observational limits on surface temperatures of nine
NSs compared with cooling curves
of NSs with several masses.
The curves are calculated
adopting proton superfluidity {\bf 1p} and neutron superfluidity 
{\bf 2nt}
in the NS cores.
}
\end{figure}

\begin{figure}
\centering
\epsfxsize=86mm
\epsffile[20 143 565 690]{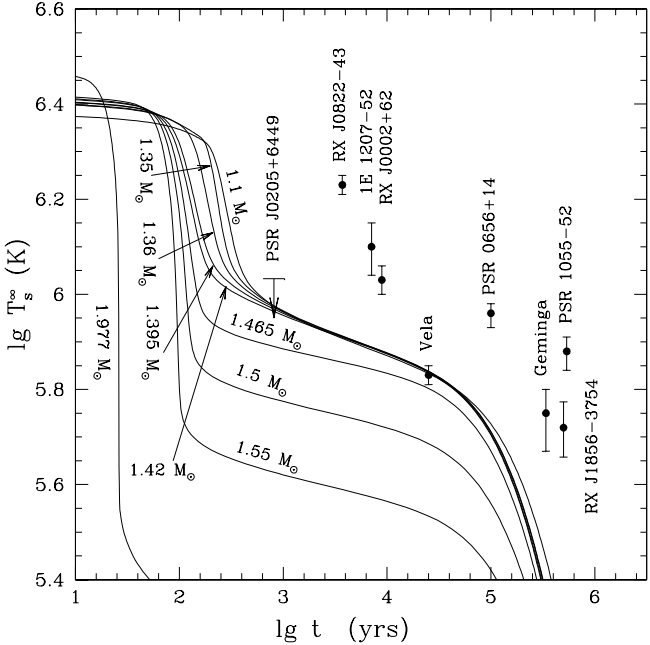}
\caption{
Same as in Fig.\ 1 but for
model {\bf 3nt} of neutron superfluidity instead of {\bf 2nt}.
}
\end{figure}

\begin{figure}
\centering
\epsfysize=6.7cm
\epsffile[20 150 570 695]{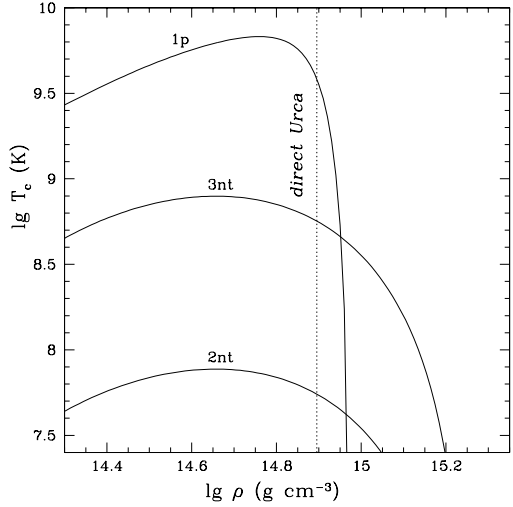}
\caption{
Density dependence of the critical temperatures
of superfluidity of protons (model {\bf 1p}) 
and neutrons (models {\bf 2nt} and {\bf 3nt}) in a NS core.
Vertical
dotted line shows direct Urca threshold, $\rho=\rho_{\rm D}$.
}
\end{figure}

\section{Cooling theory}
\label{sect2}

We confront the observational data with our simulations
of NS cooling, using the recent cooling
theory summarized in KYG and YGKP. For simplicity,
we consider the models of NSs with the cores
composed of neutrons, protons, and electrons (npe matter).
We use the equation of state (EOS) in the NS core
proposed by Prakash et al.\  (\cite{pal88})
(version I of the symmetry energy, with the compression modulus
$K=240$ MeV of the saturated nuclear matter; it is denoted as
EOS A in KYG and YGKP). The maximum NS mass for this
EOS is $M_{\rm max}= 1.977$ M$_\odot$ 
(with the central density $\rho_{\rm c}^{\rm max}=
2.575 \times 10^{15}$ g cm$^{-3}$).
The adopted EOS opens direct Urca process in the NSs with masses
$M > M_{\rm D} =1.358$ M$_\odot$ and central densities 
above $\rho_{\rm D} =7.851 \times 10^{14}$ g cm$^{-3}$.

In our simulations, we take into account superfluidity
of nucleons in the NS interiors.
Superfluidity suppresses many neutrino emission processes
(e.g., direct and modified Urca processes, 
nucleon-nucleon bremsstrahlung)
but opens a specific powerful mechanism of neutrino emission
due to the Cooper pairing of nucleons
(proposed by Flowers et al.\ \cite{frs76}; see, e.g.,
Yakovlev et al.\ \cite{ykgh01} for details),
and also affects the heat capacity of matter.
We include the singlet-state
pairing of protons and the triplet-state pairing of neutrons
in the NS cores but,
for simplicity, we neglect the singlet-state pairing of neutrons
in the NS crusts. 
The core superfluids are characterized
by the density-dependent critical temperatures $T_{\rm cp}(\rho)$
and $T_{\rm cnt}(\rho)$ (Fig.\ 3). We use one model of strong
proton superfluidity (model {\bf 1p} described, e.g., in KYG), 
and two models of triplet-state
neutron superfluidity (model {\bf 2nt} of weak superfluidity 
and model {\bf 3nt} of moderately strong superfluidity).
The critical temperatures $T_{\rm c}(\rho)$ are parameterized
by Eq.\ (1) in KYG. The parameters of model {\bf 1p}
are given in KYG; the parameters of models
{\bf 2nt} and {\bf 3nt} are the same as for model {\bf 1nt}
in KYG, but the parameter $T_0$ is now equal
to $2 \times 10^9$ K and $1.5 \times 10^{10}$ K, respectively.
Our phenomenological superfluid models
are consistent with the current
microscopic models of nucleon superfluidity in NS cores
(e.g., Lombardo \& Schulze \cite{ls01}).

\section{Theory and observations}
\label{sect3}

Figures 1 and 2 compare the observational data with theoretical
cooling curves, $T_{\rm s}^\infty(t)$. 

Adopting model {\bf 1p}
of proton superfluidity and model {\bf 2nt} of neutron superfluidity
we obtain (Fig.\ 1) a family of cooling curves 
for NSs with different masses $M$. Actually, superfluidity
{\bf 2nt} is rather weak and has almost no effect on NS cooling.
The properties of such cooling models are discussed
in KYG and YGKP. One can distinguish NSs
of three types:\\ 
(I) Low-mass NSs, $M \la M_{\rm I}$, are 
very slowly
cooling NSs where modified or direct Urca processes
are strongly suppressed by proton superfluidity; 
their cooling curves are almost independent of NS mass and EOS.\\
(II) Medium-mass NSs, $M_{\rm I} \la M \la M_{\rm II}$,
undergo moderately fast cooling via direct Urca process
partly reduced by proton superfluidity; their cooling
is very sensitive to NS mass, EOS, and $T_{\rm cp}(\rho)$
model. \\
(III) Massive NSs, $M \ga M_{\rm II}$, show
fast cooling via direct Urca process in the NS centers almost unaffected
by proton superfluidity.
At $t \sim 10^5$ yr, for our NS models,
we have $M_{\rm I} \sim 1.36$ M$_\odot$
and $M_{\rm II} \sim 1.52$ M$_\odot$.
These values are easily varied by choosing
other EOSs and proton superfluid models (KGY, YGKP).

The situation would be drastically different if
we adopted neutron superfluidity {\bf 3nt} instead of
{\bf 2nt}. We would get a number of cooling curves
plotted in Fig.\ 2. As long as a NS is
hot and its internal temperature is larger than
the maximum of $T_{\rm cnt}(\rho)$,
the neutron superfluidity is absent and the star
cools as shown in Fig.\ 1.
However, the appearance of a moderately strong
neutron superfluidity induces powerful neutrino
emission due to the Cooper pairing of neutrons,
which leads to a very fast
cooling. In low-mass NSs ($M \leq M_{\rm D}$),
where direct Urca process is forbidden,
this fast cooling has nothing to do with direct Urca process.
As seen from Fig.\ 2, one can easily explain the upper
limit of $T_{\rm s}^\infty$ for PSR J0205
by cooling of such a star. Moreover, by 
changing the
maximum of $T_{\rm cn}(\rho)$, one can explain
all relatively cool sources in Figs.\ 1 and 2
(including the coldest ones such as Vela and Geminga) by cooling
of low-mass NSs with their
own models of neutron superfluidity in the NS cores.
In this way, it seems that the current observational data do not
require direct Urca process (or similar
processes in pion or kaon condensed matter, or in quark matter).

However, the main point is that NSs
may have different masses, surface magnetic fields, etc.,
but they {\it must have} the same EOS and superfluid properties
of their cores. Thus, all the sources should be explained by
one set of models of $T_{\rm cn}(\rho)$ and $T_{\rm cp}(\rho)$.
A natural explanation (KYG, YGKP) is to assume
a weak neutron superfluidity in the NS cores (e.g., model {\bf 2nt},
Fig.\ 1) and the presence of direct
Urca process in massive NSs. If this is true,
the two hottest sources for their ages, RX J0822 and PSR 1055,
can be treated as low-mass
NSs of type I, while 1E 1207, RX J002, Vela, PSR 0656, 
Geminga, and RX J1856
can be treated as medium-mass NSs of type II.
This interpretation would be impossible
without introducing the direct Urca process. 
Notice that the revised age of RX J1856 (Walter \& Lattimer \cite{wl02})
changes its status from a type I NS (e.g., KYG)
to a type II NS. 
If PSR J0205
has the surface temperature just below the inferred
upper limit, it belongs to the family of type II NSs and requires
direct Urca process in its core. The appropriate
cooling curve 
(e.g., the $M=1.42$ M$_\odot$ curve in Fig.\ 1)
would lie {\it above} the cooling curves
for Vela and Geminga which means that  
Vela and Geminga would be {\it colder} for their ages
than PSR J0205. In other words, the well-known
observational data on Vela and Geminga 
(e.g., Pavlov et al.\ \cite{pavlovetal01},
Halpern \& Wang \cite{hw97}) provide stronger arguments in favor of direct
Urca process than the newly reported data on PSR J0205.
Let us recall that
our interpretation enables one to measure the masses of type II
NSs for a fixed EOS and superfluid properties of NS
interiors (see KYG and YGKP).
In the above scenario (Fig.\ 1), the mass of PSR J0205 would be
lower than the masses of Vela and Geminga.

\section{Conclusions}
\label{sect5}

We propose a theoretical interpretation
of the upper limit on $T_{\rm s}^\infty$ of PSR J0205+6449
reported recently by Slane et al.\ (\cite{shm02}).
Although our interpretation is based on the specific
NS models with given EOS and superfluid properties
of NS interiors, it is, in fact, quite generic. 
As discussed
in KYG and YGKP, we could arrive to qualitatively similar
conclusions by choosing other EOSs and density profiles
of the superfluid critical temperatures.
Let us emphasize that the 
current observational data are explained by cooling of
NSs with the cores composed of neutrons, protons, and electrons,
without invoking any exotic matter.
Nucleon superfluidity in NS cores is a widely
accepted phenomenon; pairing of nucleons
in atomic nuclei is proved experimentally.  On the contrary,
exotic phases of   matter (pion and kaon condensates,
quark matter) remain undetected. It is therefore
comfortable to find that the npe matter with superfluid
nucleons, which can be treated as a {\it minimal model}
of NS cores, is sufficient to explain the present observations
of cooling NSs.

Our main conclusion is that the detected upper limit
of $T_{\rm s}^\infty$, by itself, does not indicate
that direct Urca process or similar processes of
strong neutrino emission operate in the NS core.
Combined with the observational data for 
other isolated middle-aged NSs,
it gives some evidence for the direct Urca process,
although less stringent than the evidence
provided by the Vela and Geminga pulsars.

Future observations of PSR J0205+6449 
would be extremely important. If the temperature
$T_{\rm s}^\infty$
appears to be lower than 0.8 MK, it would mean
(in the frame of our interpretation) that PSR J0205+6449
is indeed the coldest observed NS. It would then provide the
strongest argument that
direct Urca process is open in the NS cores.
Were the detected temperature or the upper limit 
be around 0.3 MK, this would be a strong indication 
that PSR J0205+6449 is a rapidly cooling massive NS of type III
(no NS of such type has been observed so far).
The values of $T_{\rm s}^\infty$ below 0.15 MK 
(the lowest $T_{\rm s}^\infty$ given by the maximum-mass model)
for PSR J0205+6449 could not be explained by
the proposed theory.

\begin{acknowledgements}
We are grateful to the referee, Dany Page,
for a useful suggestion that improved the presentation of figures.
The work was partly supported by RFBR grants No.\ 02-02-17668
and 00-07-90183, and KBN grant No.\ 5P03D.020.20.
\end{acknowledgements}

\end{document}